\documentstyle[twoside,espcrc2,fleqn,epsf]{article}

\newcommand{\AmS}{{\protect\the\textfont2
  A\kern-.1667em\lower.5ex\hbox{M}\kern-.125emS}}
\newcommand{\msbar}{\overline{MS}}

\newcommand{\vecx}{{\bf x}}       
\newcommand{\vecy}{{\bf y}}       
\newcommand{\vecp}{{\bf p}}       

\newcommand{\al}{\alpha}
\newcommand{\bt}{\beta}
\newcommand{\gm}{\gamma}
\newcommand{\dl}{\delta}

\newcommand{\et}{\eta}
\newcommand{\kp}{\kappa}
\newcommand{\lm}{\lambda}
\newcommand{\rh}{\rho}

\newcommand{\ta}{\tau}

\newcommand{\ph}{\phi}
\newcommand{\vr}{\varphi}

\newcommand{\om}{\omega}

\newcommand{\Gm}{\Gamma}

\newcommand{\half}{\frac{1}{2}}

\newcommand{\Tr}{\mbox{Tr}\,}

\newcommand{\dnot}{\partial_{0}}

\newcommand{\phd}{\ph^{\dagger}}

\newcommand{\eela}[1]{\label{#1}\end{equation}}
\newcommand{\eeala}[1]{\label{#1}\end{eqnarray}}
\newcommand{\be}{\begin{equation}}
\newcommand{\ee}{\end{equation}}
\newcommand{\bea}{\begin{eqnarray}}
\newcommand{\eea}{\end{eqnarray}}

\title{Real-time simulations and the electroweak phase transition}

\author{Jan Smit\address{
Institute of Theoretical Physics, University of Amsterdam,\\
Valckenierstraat 65, 1018 XE Amsterdam, the Netherlands}
        \thanks{Supported by FOM}
        }

\begin{document}

\begin{abstract}
We review recent developments in real-time simulations of SU(2)-Higgs
theory near the electroweak transition and related topics.
\end{abstract}

\maketitle

\section{Nonequilibrium phenomena}
Nonequilibrium quantum field theory is important in various physical
situations, such as domain formation during cosmological 
phase transitions and the properties of the electroweak transition
in the early universe, or the QCD transition in heavy ion collisions.
We concentrate in this talk on the SU(2)-Higgs model
and the electroweak transition, which is relevant 
to theories of baryogenesis \cite{RuSha96}.

The description of nonequilibrium phenomena involves real time, 
as opposed to the imaginary time which is so useful for Monte Carlo 
computations in quantum field theory. Consider the time dependence 
of an observable $O$,
\[ 
\langle O(t) \rangle = \Tr \rh \, e^{iHt}\, O\, e^{-iHt}.
\] 
We know how to turn this operator expression into path integral 
form. It leads to phase factors $\exp(\pm iS)$, which are 
hopeless for Monte Carlo. Very hard is also the anaytical 
continuation of an imaginary time expression 
back to real time when the data for $\langle O(-i\ta)\rangle$ have errors, 
especially at `large' times.
A possible way out of these difficulties is the classical 
approximation.

\section{Classical SU(2)-Higgs model on a spatial lattice}

The action 
$S = \int dt\, (K - W)$, is given schematically by 
the kinetic energy 
\bea
K &=& a\sum_{\vecx}\left[
\frac{1}{z_E g^2}\, 
\Tr (D_0 U_{m})^{\dagger} D_0 U_{m} 
\right. \nonumber \\ && \mbox{} \left.
+ \frac{1}{z_{\pi}}\, 
((\dnot -iA_{0})\vr)^{\dagger} 
 (\dnot -iA_{0})\vr
\right]_{\vecx}
\nonumber
\eea
and the potential energy
\bea
W &=& \frac{1}{a}\sum_{\vecx}
\left[ \sum_{mn}\frac{1}{g^2}\,\Tr(1-U_{mn})
\right.\label{HW} \\ && \left. \mbox{}
+ (D_m\vr)^{\dagger} D_m\vr
+ \mu^2 \vr^{\dagger} \vr + \lm (\vr^{\dagger} \vr)^2
\right]_{\vecx}.
\nonumber
\eea
Here time and $A_0$ are in physical units, but everything else is in
lattice units $a=1$, with the exception of the explicitly indicated
lattice spacing $a$.

Starting from an arbitrary initial field configuration the system 
will evolve in time over a region in phase space compatible with 
the conserved quantities. By ergodicity the field configurations 
sampled at large time intervals will be distributed according to 
the microcanonical ensemble. For large systems this is equivalent 
to the canonical ensemble: time average $\approx$ canonical 
average. Such averaging implies thermal equilibrium but we can 
still study small departures from equilibrium by looking at 
linear response functions $\langle O(t) O'(0) \rangle$. Once the 
properties of these quantities are well understood we can turn to 
larger departures from equilibrium.

In the canonical description in the temporal gauge $A_0 = 0$, 
Gauss' law $\dl S/\dl A_{0\vecx}^{\al} \equiv G_{\vecx}^{\al} = 0$
has the status of a constraint.
The hamiltonian takes the form $H = K + W$, with now
\[
K = \frac{1}{a} \sum_{\vecx} \left[\frac{z_E}{2}\, g^2
E^{\al}_{m} E^{\al}_{m}
+ z_{\pi} \pi^{\dagger} \pi\right]_{\vecx},
\]
and the canonical partition function
\[
Z = \int DE D\pi DU D\vr\, [\prod_{\vecx\al} 
\dl(G_{\vecx}^{\al})] \, \exp(-H/T).
\]
By equipartition any initial configuration which is smooth on the 
lattice scale evolves into a `rough' configuration, with 
temperature given by
\[ 
\frac{\langle K \rangle}{V} = \half\, (3 \times 3 + 4 - 3)\, 
\frac{T}{a^3},
\] 
where $V$ is the volume and $\mbox{}-3$ reflects the three Gauss 
contraints per lattice site $\vecx$. The energy density 
diverges as the lattice spacing $a \to 0$ at fixed temperature, 
which is the notorious Rayleigh-Einstein-Jeans divergence \cite{Pais}.

\section{Relation to dimensional reduction}
The REJ divergence is not 
the only one, which can be seen by studying time-independent
correlation functions like 
$\langle Q_{\vecx}(t) O_{\vecy}(t) \rangle$.
It is then illuminating to integrate out the 
momenta by reintroducing $A_0$ into the canonical partition 
function, using
$\int \exp(iA_{0\vecx}^{\al} G_{\vecx}^{\al}) dA_{0\vecx}^{\al}$
to represent $\dl(G_{\vecx}^{\al})$. 
This leads via a rescaling of 
$A_0$ to the form of a three dimensional euclidean
field  theory,
\[ 
Z_{\rm DR} = \int DA_0 DU D\vr \exp(-S_{\rm DR}),
\] 
with
\bea
S_{\rm DR} &=&
\frac{W}{T} + \frac{1}{g^2 aT} \sum_{\vecx} \left[\half\, D_m 
A_{0}^{\al} D_m A_{0}^{\al} 
\right. \nonumber \\ && \left. \mbox{}
+ \frac{z_E/z_{\pi}}{4}\, 
g^2 \vr^{\dagger}\vr A_{0}^{\al} A_{0}^{\al}\right]_{\vecx}.
\label{SDR}
\eea
Hence, for static quantities the classical theory is a dimensional
reduction approximation to the quantum theory.

Dimensional reduction is an accurate approximation to a weakly coupled
quantum theory in equillibrium at high temperature \cite{Ja96Rumm97}.
The reduced 3D theory is superrenormalizable which
means that only mass counterterms are needed to obtain a finite theory.
However, the above $S_{\rm DR}$ lacks a mass term for the `adjoint Higgs
field' $A_0$, since this is prohibited by the locality and
gauge invariance of the
classical action. In contrast, the usual derivation of the DR theory
leads to additional terms of the form
$\mu_A^2 \Tr A_0^2$ and $\lm_A \Tr A_0^4$.
The coefficient $\lm_A$ turns out to be negligible, but the fact that
$\mu_A \equiv 0$ in the above $S_{\rm DR}$ means that the there is 
no possibility for an $A_0^2$  mass counterterm. It follows
that the Debije screening
mass $m_D$ associated with $A_0$ is divergent in perturbation theory,
$m_D \propto 1/\sqrt{a}$.
This need not be a disaster as $a \to 0$, because $A_0$ will simply
decouple as it gets heavy.
Indeed, $A_0$ is often integrated out explicitly as an additional
approximation to dimensional reduction, because the renormalized
Debije mass is large in perturbation theory.

The classical action evidently has to be interpreted as an effective action. 
Its parameters can be
found by comparison with the dimensionally reduced quantum action
with $A_0$ integrated out,
which is known analytically through perturbative
calculations. This means that $g^2$, $\lm$ and $\mu^2$ are explicitly
known in terms of the corresponding parameters of the quantum theory, 
and $a$ and $T$ \cite{TaSm96}. 
In fact, $g^2 \approx g^2_{\msbar}(7T)$.
This comparison also shows that the ratio 
$z_E/z_{\pi} \approx 1$ to a very good approximation. 

We conclude, that for static observables the classical theory 
can approximate the quantum theory well.
There is one undetermined parameter $z = z_E =z_{\pi}$.
This parameter sets the time scale relative to 
the lattice spacing and it therefore
plays a role in dynamical quantities, to which we turn next.

\section{Dynamics}

Time dependent quantities like $\langle O(t) O(0) \rangle$ can be
computed in the microcanonical ensemble by solving the equations
of motion on a computer. We can `help ergodicity' 
by using the canonical ensemble for generating many initial 
configurations and averaging over these.
Luckily, there are now two good solutions to the algorithmic problem of the 
implementation of the Gauss constraint \cite{Kra95,Mo96}.

To see how well the classical theory may fare for dynamical quantities,
one may turn to pertubation theory.
The problem of solving perturbatively the equations of motion and
averaging over initial conditions with the canonical ensemble has been
studied in \cite{BoMcLeSmi,AaSm96} for $\ph^4$ scalar field theory.
It was concluded in \cite{AaSm96}
that the mass counterterm needed to make static correlation functions finite is
also sufficient for obtaining finite time-dependent correlation functions. 
The classical $\ph^4$
theory is renormalizable in this sense. Furthermore, after matching
parameters of the classical theory to the dimensionally reduced
quantum theory, the classical plasmon damping rate turned out to be
identical to the quantum rate to leading order in the temperature and
coupling \cite{AaSm96} (see also \cite{BuJa97}). 
Further analysis led to the conclusion that the classical $\lm\ph^4$ theory
can approximate the quantum theory for momenta and frequencies 
up to $O(\sqrt{\lm} T)$, with corrections $O(\sqrt{\lm})$ \cite{AaSm97}.

The procedure of solving equations of motion and averaging over initial
conditions is awkward analytically.
A more convenient formalism can be given 
which is a classical analogue 
of thermal field theory \cite{AaSm97}. Alternatively one
may use the imaginary time formalism of
finite temperature quantum field theory
and replace (after analytic continuation to real time)
the Bose distribution by its high temperature approximation: 
\be
\half\, + \frac{1}{e^{\om/T} -1} \to \frac{T}{\om}.
\label{quatocla}
\ee
We assume this
latter procedure for a brief discussion of the situation in gauge theories.

\begin{figure}[t]
\epsfxsize=75mm 
\centerline{\epsfbox{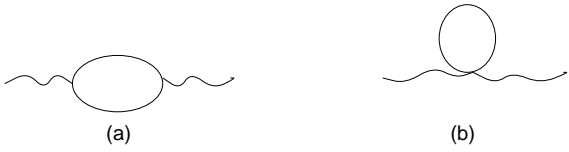}}
\caption{
Gauge boson self energy diagrams.
}
\label{f1}
\end{figure}

Consider the gauge boson selfenergy\\ $\Pi_{\mu\nu}(\vecp,p^0)
= \Pi_{\mu\nu}^{(a)}(\vecp,p^0) + \Pi_{\mu\nu}^{(b)}(\vecp,p^0)$, 
as given by the diagrams in fig.\ \ref{f1},
using the abelian Higgs model as a simple example. Both contributions
(a) and (b) are linearly divergent (the quadratic divergence of the
quantum theory is reduced to a linear divergence in the classical theory
by (\ref{quatocla})). Because of gauge invariance we may
expect cancelations. Indeed, for external frequency  
$p^0= 0$ the fields are static and therefore the
spatial components $\Pi_{mn}^{(a)}(\vecp,0)$ and $\Pi_{mn}^{(b)}(\vecp,0)$
are related by a Ward identity of the dimensionally reduced theory.
As a consequence, their sum $\Pi_{mn}(\vecp,0)$ is finite.
However, for $p^0 \neq 0$, the cancellation is incomplete and
$\Pi_{mn}(\vecp,p^0)$ turns out to be linearly divergent, in both real 
and imaginary parts. This may seem unfamiliar because in the quantum
theory at zero temperature the Ward indentity reduces 
a quadratic divergence into a logarithmic
divergence for all momenta. The reason such reduction does not follow here
is that at finite temperature
$\Pi_{mn}(\vecp,p^0)$ is not analytic at $\vecp = 0$
for $p^0 \neq 0$. This is related to 
the physical process of Landau damping and a detailed physical picture has been
developed by Arnold \cite{Ar97} which is also valid in the nonabelian case 
(on the lattice,
measure-effects end up only into the momentum independent $\Pi_{\mu\nu}^{(b)}$).

The momentum dependence of the divergent terms is complicated, typically of the
hard thermal loop form $\ln[(p^0 + |\vecp|)/(p^0 - |\vecp|)]$.  This means that
the divergencies in classical
gauge theory are nonlocal in spacetime (in contrast to scalar
field theory), which suggests nonlocal counterterms. This does
not look attractive, although it may be possible to introduce new degrees of
freedom to re-express such counterterms in local form. 
The situation is complicated, however, by the fact
that on the lattice the divergencies lack rotational covariance 
\cite{BoMcLeSmi,Ar97}. 

A different point of view prevails, in which the lattice spacing is supposed
to stay finite, of order of the inverse temperature
(the linear divergence $\propto g^2 T/a$ is cured to $\propto g^2 T^2$ in the 
quantum theory), and where the hard thermal
loop physics is to be added `by hand'. There are difficulties with
double counting, but practical proposals \cite{HuMu} are being pursued.
Yet another approach consists of deriving an effective theory in which 
high frequency modes are subdominant, such that regulator effects vanish as the 
lattice spacing goes to zero \cite{Son}.

In conclusion, for time-dependent quantities the classical model suffers from
hard thermal loop effects which have lattice artefacts and which are 
divergent as the lattice spacing goes to zero. 
This is also relevant for Lyapunov exponents 
(cf.\ the brief remarks in \cite{TaSm96}). 
Awaiting a solution of these
problems, we still may expect to obtain useful information with 
the classical model at finite lattice spacing, provided the results
are interpreted with care \cite{Ar97}. 

\section{Sphaleron rate}

The sphaleron rate, plays an important role in theories of baryogenesis
\cite{RuSha96}. It is the real time analogue of the topological susceptibility
in imaginary time,
\be
\Gm = \frac{1}{t}\,\langle \left[
\int_0^t dx^0 \int d^3 x\, \frac{\Tr F \tilde F}{16\pi^2}
\right]^2 \rangle,
\;\;\; t \to \infty,
\label{defrate}
\ee
in continuum notation. The usual expectations have been
\bea
\kp &\equiv& \frac{\Gm/V}{(\al_W T)^4},
\;\;\; \al_W = \frac{g^2}{4\pi}, \label{defkap}\\
&\propto& \exp[- E_s(T)/T],\;\;\; T < T_c, \label{lowT}\\
&\approx& \mbox{const}, \;\;\;\;\;\;\;\;\;\;\;\;\;\;\;\;\;\;\;    T > T_c,
\label{highT}
\eea
where 
$E_s (T)$ is the temperature dependent sphaleron energy (of order
10 TeV for $T=0$, vanishing near $T_c$)
and $\mbox{const} = O(1)$. The prefactors of the exponential sphaleron
suppression (\ref{lowT}) have been calculated analytically
and the main problem for the numerical simulations is the computation
of $\kp$ above $T_c$. Recent work in pure SU(2) gauge theory \cite{AmKra95}
and in SU(2)-Higgs theory \cite{TaSm96}
using the same numerical implementation
found $\kp \approx 1.1$ in
the high temperature phase. Furthermore, the results did not appear to depend
on the lattice spacing. This is quite remarkable: 
a reduction of $a$ by a factor 0.6 caused $\Gm/V$ to fall by factor 
$(0.6)^4 \approx 0.13$, in both phases \cite{TaSm96}. 
However, the expected
sphaleron suppression was not observed in the low temperature phase,
and indeed the results are considered to be wrong. The reason
is that the imperfections of the `naive'
lattice implementation used for $\Tr F \tilde F$ imply not only the
need for a multiplicative renormalization, but also a
subtractive renormalization for the rate; these were not taken into account. 
Such a subtraction has been controversial in euclidean lattice QCD, but in 
practise it appears to work well \cite{DiGia}.

Recently, there have been
two new computations using improved
methods for obtaining $\Gm$ \cite{MoTu97,AmKra97,Kra97}, 
but before presenting these it is useful to mention new
theoretical analysis questioning the lorical $\al_W$ independence
of `const' in (\ref{highT}).
Arnold, Son and Yaffe (ASY) proposed the following picture \cite{ArSoYa97}:

the nonperturbative processes important for $\Gm$
occur on the typical momentum scale
$g^2 T$; however, due to Landau damping,
the corresponding frequencies are not of the order $g^2 T$,
but $g^4 T$.

Hence, the prediction is 
$\Gm/V \approx \mbox{const}'\, g^2\, (g^2 T)^4$ or
$\kp \propto g^2$ as $g^2 \to 0$, in contrast to
the lore that $\kp$ in (\ref{highT}) is practically independent of $g^2$.

Now
$g^2 T$ is the natural classical (dimensional reduction) scale for
static processes ($H/T = \int d^3 x\, \Tr F_{mn} F_{mn} /4g^2 T + \cdots$), 
but the extra factor of $g^2$ can only be supplied in a given regularization.
On the lattice this involves the combination $\bt \equiv 4/g^2 aT$ 
(recall (\ref{HW})).
So the prediction is that $\kp \propto \bt^{-1}$, i.e.\ 
$\propto a$! 
Physically, the damping in the classical theory diverges as $a \to 0$,
as discussed in the previous section, and then nothing moves anymore. 

A detailed study of lattice effects in the classical theory by Arnold 
\cite{Ar97}
suggests that not all is lost: we should readjust the time scale in the
classical theory proportional to the ratio of damping rates of the 
quantum and classical theory. He obtained the following matching relation:
\be
\kp \approx 3.2\, \al_W\, \bt\, \kp_{\rm class},
\label{formula}
\ee
with an estimated error of about 30 \% due to the cubic anisotropy of 
the lattice.

\begin{figure}[t]
\epsfxsize=75mm 
\centerline{\epsfbox{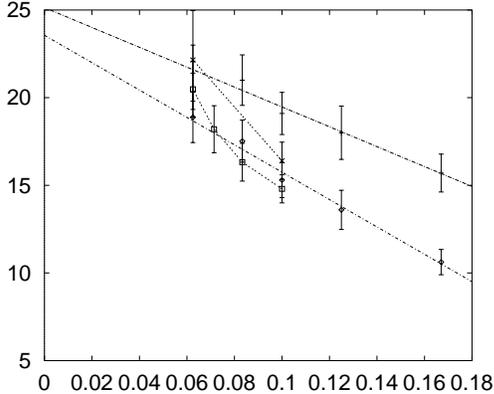}}
\caption{
Data for $\bt\kp_{\rm class}$ versus $\bt^{-1}$. 
The straight lines are fits to
the MT data \cite{MoTu97}: 
$\kp_{\rm class} = 25.1\, \bt^{-1} - 56.7\, \bt^{-2}$ (upper)
and $\kp_{\rm class} = 23.6\, \bt^{-1} - 78.1\, \bt^{-2}$ (lower). 
The AK data are connected by lines to guide the eye; 
lower: ref.\ \cite{AmKra97}, upper: ref.\ \cite{Kra97}.
}
\label{f2}
\vspace{-0.6cm}
\end{figure}

The new methods for the computation of $\Gm$ use cooling 
(Ambj\o rn and Krasnitz (AK) \cite{AmKra97,Kra97})
or computing winding numbers of gauge transformations 
(Moore and Turok (MT) \cite{MoTu97}).
With either method the rate in the low temperature phase was found too small
to be measureable, which is what one expects because of
the sphaleron suppression (\ref{lowT}).
Fig.\ \ref{f2} summarizes the results for $\kp_{\rm class}$ 
in the high temperature
phase, in the form of $\bt\kp_{\rm class}$ as a function of $\bt^{-1}$. 
The lower MT data are perturbatively corrected
for some discretization errors. The 
AK data should be compared with the upper MT data, and they can
be seen to be compatible, but the trend as a function of
$\bt$ is different.
If
$\kp_{\rm class}$ indeed vanishes proportional to $a$ we expect 
$\bt\kp_{\rm class}$ to approach  
a constant as $\bt^{-1} \to 0$. This is clearly suggested by the MT data
which can be fitted very well by a straight line
(the slope represents $O(a^2)$ corrections), 
but not by the AK data, for which $\kp_{\rm class}$ itself and not 
$\bt\kp_{\rm class}$
is approximately constant. 
Given the much larger leverage in the MT
data one may tentatively conclude
that $\kp_{\rm class} \sim \bt^{-1}$. 
Of course, smaller $\bt^{-1}$ are desirable to confirm this behavior,
or not.  However, small volume AK results also show 
$\kp_{\rm class} \propto \bt^{-1}$ \cite{AmKra97}. 
Using (\ref{formula}) and the corrected MT data
the tentative conclusion is then $\kp \approx 76\, \al_W$.

\begin{figure}[t]
\epsfxsize=75mm
\centerline{\epsfbox{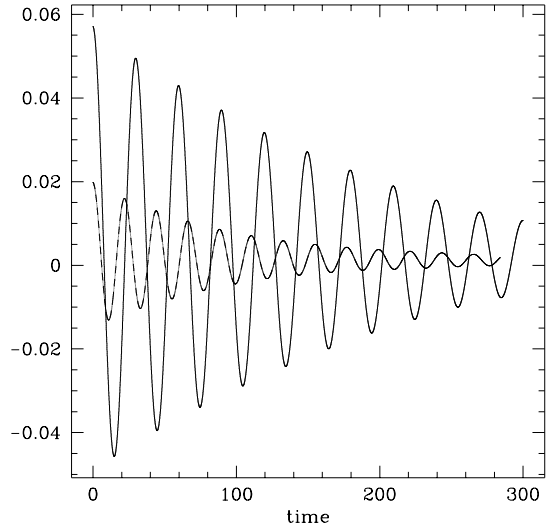}}
\caption{The autocorrelator $C_H(t)$ in the Higgs phase for
$T/T_c = 0.85$, $aT = 0.96$ (small amplitude) and $T/T_c = 0.78$, 
$aT = 0.41$ (large amplitude). The time is in lattice units.
}
\label{f3}
\vspace{-0.6cm}
\end{figure}
\begin{figure}[t]
\epsfxsize=75mm
\centerline{\epsfbox{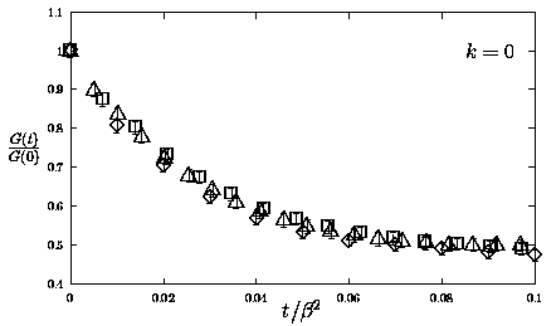}}
\caption{Zero momentum gauge field
correlation function in Coulomb gauge, $C_A(t)/C_A(0)$ 
versus $t/\beta^2$, for $\bt = 10$, 12, 14; ref.\ \cite{AmKra97}.
}
\label{fkra0}
\vspace{-1.0cm}
\end{figure}

\section{Autocorrelation functions}
Apart from the correlation (\ref{defrate}) used for the computation of the
sphaleron rate there is very little nonperturbative information on 
time-dependent correlation functions. Correlators  
$C_{O} = \langle O(t) O(0) \rangle$ of gauge invariant fields
\[
O \to \vr^{\dagger} \vr \equiv H,
\;\;\;
\to i\Tr \phd (\partial_k - i A_k) \ph \tau_{\al} \equiv W^{\al}_k,
\]
($\ph$ is the matrix version of the doublet $\vr$)
have been computed at zero momentum \cite{TaSm97}.
%
Their interpretation may be guided by perturbation theory.
In the Higgs phase $H$ and $W_k^{\al}$ are equivalent to $\vr$ and $A_k^{\al}$,
to lowest order, but in the plasma (high temperature) phase the interpretation
is not so straightforward. In particular the analogy with the confinement
phase of QCD, which holds for time-independent correlation functions and which
suggests pole dominance by confined states,
appears to be misleading in this time-dependent case.
Furthermore, there is to my knowledge no analogue of the zero temperature
`theorem of the arbitrariness of the interpolating field'. 

Fig.\ \ref{f3} shows an example of $C_H$ in the Higgs phase. Similar results
were obtained for $C_W(t)$. 
From the 
damped oscillations `plasmon' frequencies and damping rates $\om$ and
$\gm$ were extracted by fitting to a pole dominance form 
$C(t) = R\exp(-\gm t)\cos(\om t + \al) + \mbox{background}$.
In the plasma phase the correlators are noisier; pole dominance is
questionable and the extraction of $\gm$ is difficult especially for $H$.
The resulting $\om$ and $\gm$ appear to be approximately independent of
the lattice spacing. Closer examination in the Higgs phase reveals, however, 
that the data for $\om_W$ are compatible with a weak lattice spacing dependence 
expected from the perturbative $1/\sqrt{a}$ divergence. 
On the other hand, the damping rates are not divergent to leading order and
both $\gm_H$ and $\gm_W$ turn out to have magnitudes similar to the 
analytic results in the quantum theory.

Ambj\o rn and Krasnitz \cite{AmKra97} studied correlators of the
gauge field itself in Coulomb gauge, $O \to A_k^{\al}$, 
for various momenta $\vecp$, in pure SU(2) gauge theory.
These are better suited for comparison with perturbation theory and
currently a hot issue is if the ASY analysis \cite{Ar97,ArSoYa97} applies.
This suggests overdamped correlation functions (which
do not oscillate), while the classical divergence will show up
in time scales increasing $\propto \bt$, i.e.\ in lattice units
$\propto \bt^2$. 
The analysis applies to a regime $p_0^2 \ll \vecp^2 \ll T^2$,
which is hard to satisfy in current simulations. Surprisingly
the characteristics of overdamping and lattice time scale 
$\propto \bt^2$
shows up even for $\vecp = 0$, cf.\ Fig.\ \ref{fkra0}. 
This remains to be explained.
 
We have not shown here the richer structures found 
in the correlation functions in the plasma phase \cite{TaSm97,AmKra97},
which are currently being analyzed \cite{BoLa97,ArYa97}.

\section{Dynamics of the phase transition}
In a very stimulating paper \cite{MoTu96},
Moore and Turok studied the real time
properties of the electroweak phase transition by numerical simulation.
They obtained 
the drag coefficient for the moving wall between the Higgs and plasma phase,
$\et = \mbox{pressure}/\mbox{wall velocity}$, 
from the fluctuations of the wall position 
via a fluctation-dissipation argument. Another computation was
the change in $\int F\tilde F$ under influence of a chemical potential
related fermion production, in moving walls. There were several other
interesting results and techniques,
but there is no more space here for an illustration. 

I thank Gert Aarts, Peter Arnold and Alex Krasnitz for
useful discussions.


\begin{thebibliography}{9}
\bibitem{RuSha96}   V.A.~Rubakov, M.A.~Shaposhnikov, hep-ph/9603208.
\bibitem{Pais}      A.\ Pais, ``Subtle is the Lord ...'', sect.\ 19b.
                    Oxford Unversity Press 1982.
\bibitem{Ja96Rumm97}
                    K.~Rummukainen, Nuc.\ Phys.\ (Proc.\ Suppl.) 53 (1997) 30.
\bibitem{TaSm96}    W.H.~Tang, J.~Smit, Nucl. Phys. B482 (1996) 265.
\bibitem{Kra95}     A.~Krasnitz, Nucl.\ Phys.\ B455 (1995) 320;
\bibitem{Mo96}      G.D.~Moore, Nucl.\ Phys.\ B480 (1996) 657.
\bibitem{BoMcLeSmi} D.~B\"odeker, L.~McLerran, A.~Smilga,
                    Phys.\ Rev.\ D52 (1995) 4675.
\bibitem{AaSm96}    G.~Aarts, J.~Smit, Phys.\ Lett.\ B393 (1997) 395.
\bibitem{BuJa97}    W.~Buchm\"uller, A.~Jakov\'ac, hep-ph/9705452.
\bibitem{AaSm97}    G.~Aarts, J.~Smit, hep-ph/9707342.
\bibitem{Ar97}      P.~Arnold, Phys.\ Rev.\ D 55 (1997) 7781.
\bibitem{HuMu}      C.R.~Hu, B.~M\"uller, hep-ph/9611292.
\bibitem{Son}       D.T.~Son, hep-ph/9707351.
\bibitem{AmKra95}   J.~Ambj\o rn, A.~Krasnitz, Phys.\ Lett.\ B362 (1995) 97.
\bibitem{DiGia}     B.~All\'es, these proceedings.
\bibitem{MoTu97}    G.D.~Moore, N.G.~Turok, hep-ph/9703266.
\bibitem{AmKra97}   J.~Ambj\o rn, A.~Krasnitz, hep-ph/9705380.
\bibitem{Kra97}     A.~Krasnitz, these proceedings.
\bibitem{ArSoYa97}  P.~Arnold, D.~Son, L.G.~Yaffe,
                    Phys.\ Rev.\ D55 (1997) 6264.
\bibitem{TaSm97}    W.H.~Tang, J.~Smit, hep-lat/9702017.
\bibitem{BoLa97}    B.~B\"odeker, M.~Laine, hep-ph/9707489.
\bibitem{ArYa97}    P.~Arnold, L.G.~Yaffe, hep-ph/9709449.
\bibitem{MoTu96}    G.D.~Moore, N.~Turok, Phys.\ Rev.\ D55 (1997) 6538.

                   
\end{thebibliography}
\end{document}